\documentclass[author-year, 11pt]{elsarticle} 
\usepackage[hyphens]{url}
\biboptions{sort&compress} 
\usepackage{graphicx}
\usepackage{booktabs} 
\makeatletter
\def\ps@pprintTitle{%
 \let\@oddhead\@empty
 \let\@evenhead\@empty
 \def\@oddfoot{\it \hfill\today}%
 \let\@evenfoot\@oddfoot}
\makeatother

\usepackage[T1]{fontenc}
\usepackage{lmodern}
\usepackage{amssymb,amsmath}
\usepackage{ifxetex,ifluatex}
\usepackage{fixltx2e} 
\IfFileExists{upquote.sty}{\usepackage{upquote}}{}
\ifnum 0\ifxetex 1\fi\ifluatex 1\fi=0 
  \usepackage[utf8]{inputenc}
\else 
  \usepackage{fontspec}
  \ifxetex
    \usepackage{xltxtra,xunicode}
  \fi
  \defaultfontfeatures{Mapping=tex-text,Scale=MatchLowercase}
  
\fi
\IfFileExists{microtype.sty}{\usepackage{microtype}}{}
\usepackage{color}
\usepackage{fancyvrb}

\DefineVerbatimEnvironment{Highlighting}{Verbatim}{commandchars=\\\{\}}
\usepackage{framed}
\definecolor{shadecolor}{RGB}{248,248,248}
\newenvironment{Shaded}{\begin{snugshade}}{\end{snugshade}}
\newcommand{\KeywordTok}[1]{\textcolor[rgb]{0.13,0.29,0.53}{\textbf{{#1}}}}
\newcommand{\DataTypeTok}[1]{\textcolor[rgb]{0.13,0.29,0.53}{{#1}}}
\newcommand{\DecValTok}[1]{\textcolor[rgb]{0.00,0.00,0.81}{{#1}}}

\newcommand{\StringTok}[1]{\textcolor[rgb]{0.31,0.60,0.02}{{#1}}}

\newcommand{\NormalTok}[1]{{#1}}
\ifxetex
  \usepackage[setpagesize=false, 
              unicode=false, 
              xetex]{hyperref}
\else
  \usepackage[unicode=true]{hyperref}
\fi
\hypersetup{breaklinks=true,
            bookmarks=true,
            pdfauthor={},
            pdftitle={RNeXML: a package for reading and writing richly annotated phylogenetic, character, and trait data in R},
            colorlinks=true,
            urlcolor=blue,
            linkcolor=magenta,
            pdfborder={0 0 0}}
\begin{document}
\begin{frontmatter}

  \title{RNeXML: a package for reading and writing richly annotated phylogenetic,
character, and trait data in R}
    \author[cstar]{Carl Boettiger\corref{c1}}
   \ead{cboettig(at)gmail.com} 
   \cortext[c1]{Corresponding author}
    \author[ropensci]{Scott Chamberlain}

    \author[NBC]{Rutger Vos}

    \author[dukeplus]{Hilmar Lapp}

      \address[cstar]{Center for Stock Assessment Research, Department of Applied Math and
Statistics, University of California, Mail Stop SOE-2, Santa Cruz, CA
95064, USA}    
    \address[ropensci]{University of California, Berkeley, CA, USA}    
    \address[NBC]{Naturalis Biodiversity Center, Leiden, the Netherlands}    
    \address[dukeplus]{Center for Genomic and Computational Biology, Duke University, and
National Evolutionary Synthesis Center, Durham, NC, USA}    
  
  \begin{abstract}
  \begin{enumerate}
  \def\labelenumi{\arabic{enumi}.}
  \item
    NeXML is a powerful and extensible exchange standard recently proposed
    to better meet the expanding needs for phylogenetic data and metadata
    sharing. Here we present the RNeXML package, which provides users of
    the R programming language with easy-to-use tools for reading and
    writing NeXML documents, including rich metadata, in a way that
    interfaces seamlessly with the extensive library of phylogenetic tools
    already available in the R ecosystem.
  \item
    Wherever possible, we designed RNeXML to map NeXML document contents,
    whose arrangement is influenced by the format's XML Schema definition,
    to their most intuitive or useful representation in R. To make NeXML's
    powerful facility for recording semantically rich machine-readable
    metadata accessible to R users, we designed a functional programming
    interface to it that hides the semantic web standards leveraged by
    NeXML from R users who are unfamiliar with them.
  \item
    RNeXML can read any NeXML document that validates, and it generates
    valid NeXML documents from phylogeny and character data in various R
    representations in use. The metadata programming interface at a basic
    level aids fulfilling data documentation best practices, and at an
    advanced level preserves NeXML's nearly limitless extensibility, for
    which we provide a fully working demonstration. Furthermore, to lower
    the barriers to sharing well-documented phylogenetic data, RNeXML has
    started to integrate with taxonomic metadata augmentation services on
    the web, and with online repositories for data archiving.
  \item
    RNeXML allows R's rich ecosystem to read and write data in the NeXML
    format through an interface that is no more involved than reading or
    writing data from other, less powerful data formats. It also provides
    an interface designed to feel familiar to R programmers and to be
    consistent with recommended practices for R package development, yet
    that retains the full power for users to add their own custom data and
    metadata to the phylogenies they work with, without introducing
    potentially incompatible changes to the exchange standard.
  \end{enumerate}
  \end{abstract}
  
 \end{frontmatter}

\section{Introduction}\label{introduction}

Users of the popular statistical and mathematical computing platform R
(R Core Team 2014) enjoy a wealth of readily installable comparative
phylogenetic methods and tools (O'Meara 2014). Exploiting the
opportunities arising from this wealth for complex and integrative
comparative research questions relies on the ability to reuse and
integrate previously generated or published data and metadata. The
expanding data exchange needs of the evolutionary research community are
rapidly outpacing the capabilities of most current and widely used data
exchange standards (Vos \emph{et al.} 2012), which were all developed a
decade or more ago. This has resulted in a radiation of different data
representations and exchange standard ``flavors'' that are no longer
interoperable at the very time when the growth of available data and
methods has made that interoperability most valuable. In response to the
unmet needs for standardized data exchange in phylogenetics, a modern
XML-based exchange standard, called NeXML, has recently been developed
(Vos \emph{et al.} 2012). NeXML comprehensively supports current data
exchange needs, is predictably machine-readable, and is forward
compatible.

The exchange problem for phylogenetic data is particularly acute in
light of the challenges in finding and sharing phylogenetic data without
the otherwise common loss of most data and metadata semantics (Stoltzfus
\emph{et al.} 2012; Drew \emph{et al.} 2013; Cranston \emph{et al.}
2014). For example, the still popular NEXUS file format (Maddison
\emph{et al.} 1997) cannot consistently represent horizontal gene
transfer or ambiguity in reading a character (such as a DNA sequence
base pair). This and other limitations have led to modifications of
NEXUS in different ways for different needs, with the unfortunate result
that NEXUS files generated by one program can be incompatible with
another (Vos \emph{et al.} 2012). Without a formal grammar, software
based on NEXUS files may also make inconsistent assumptions about
tokens, quoting, or element lengths. Vos et al. (2012) estimates that as
many as 15\% of the NEXUS files in the CIPRES portal contain
unrecoverable but hard to diagnose errors.

A detailed account of how the NeXML standard addresses these and other
relevant challenges can be found in Vos et al. (2012). In brief, NeXML
was designed with the following important properties. First, NeXML is
defined by a precise grammar that can be programmatically
\textbf{validated}; i.e., it can be verified whether a file precisely
follows this grammar, and therefore whether it can be read (parsed)
without errors by software that uses the NeXML grammar (e.g.~RNeXML) is
predictable. Second, NeXML is \textbf{extensible}: a user can define
representations of new, previously unanticipated information (as we will
illustrate) without violating its defining grammar. Third and most
importantly, NeXML is rich in \textbf{computable semantics}: it is
designed for expressing metadata such that machines can understand their
meaning and make inferences from it. For example, OTUs in a tree or
character matrix for frog species can be linked to concepts in a
formally defined hierarchy of taxonomic concepts such as the Vertebrate
Taxonomy Ontology (Midford \emph{et al.} 2013), which enables a machine
to infer that a query for amphibia is to include the frog data in what
is returned. (For a more broader discussion of the value of such
capabilities for evolutionary and biodiversity science we refer the
reader to Parr et al. (2011).)

To make the capabilities of NeXML available to R users in an easy-to-use
form, and to lower the hurdles to adoption of the standard, we present
RNeXML, an R package that aims to provide easy programmatic access to
reading and writing NeXML documents, tailored for the kinds of use-cases
that will be common for users and developers of the wealth of
evolutionary analysis methods within the R ecosystem.

\section{The RNeXML package}\label{the-rnexml-package}

The \texttt{RNeXML} package is written entirely in R and available under
a Creative Commons Zero public domain waiver. The current development
version can be found on Github at
\href{}{\url{https://github.com/ropensci/RNeXML}}, and the stable
version can be installed from the CRAN repository. \texttt{RNeXML} is
part of the rOpenSci project. Users of \texttt{RNeXML} are encouraged to
submit bug reports or feature requests in the issues log on Github, or
the phylogenetics R users group list at
\texttt{r-sig-phylo@r-project.org} for help. Vignettes with more
detailed examples of specific features of RNeXML are distributed with
the R package and serve as a supplement to this manuscript. Each of the
vignettes can be found at
\href{}{\url{http://ropensci.github.io/RNeXML/}}.

\subsection{Representation of NeXML documents in
R}\label{representation-of-nexml-documents-in-r}

Conceptually, a NeXML document has the following components: (1)
phylogeny topology and branch length data, (2) character or trait data
in matrix form, (3) operational taxonomic units (OTUs), and (4)
metadata. To represent the contents of a NeXML document (currently in
memory), \texttt{RNeXML} defines the \texttt{nexml} object type. This
type therefore holds phylogenetic trees as well as character or trait
matrices, and all metadata, which is similar to the phylogenetic data
object types defined in the \texttt{phylobase} package (NESCENT R
Hackathon Team 2014), but contrasts with the more widely used ones
defined in the \texttt{ape} package (Paradis \emph{et al.} 2004), which
represents trees alone.

When reading and writing NeXML documents, \texttt{RNeXML} aims to map
their components to and from, respectively, their most widely used
representations in R. As a result, the types of objects accepted or
returned by the package's methods are the \texttt{phylo} and
\texttt{multiPhylo} objects from the \texttt{ape} package (Paradis
\emph{et al.} 2004) for phylogenies, and R's native \texttt{data.frame}
list structure for data matrices.

\subsection{Reading phylogenies and character
data}\label{reading-phylogenies-and-character-data}

The method \texttt{nexml\_read()} reads NeXML files, either from a local
file, or from a remote location via its URL, and returns an object of
type \texttt{nexml}:

\begin{Shaded}
\begin{Highlighting}[]
\NormalTok{nex <-}\StringTok{ }\KeywordTok{nexml_read}\NormalTok{(}\StringTok{"components/trees.xml"}\NormalTok{)}
\end{Highlighting}
\end{Shaded}

The method \texttt{get\_trees\_list()} can be used to extract the
phylogenies as an \texttt{ape::multiPhylo} object, which can be treated
as a list of \texttt{ape::phylo} objects:

\begin{Shaded}
\begin{Highlighting}[]
\NormalTok{phy <-}\StringTok{ }\KeywordTok{get_trees_list}\NormalTok{(nex)}
\end{Highlighting}
\end{Shaded}

The \texttt{get\_trees\_list()} method is designed for use in scripts,
providing a consistent and predictable return type regardless of the
number of phylogenies a NeXML document contains. For greater convenience
in interactive use, the method \texttt{get\_trees()} returns the R
object most intuitive given the arrangement of phylogeny data in the
source NeXML document. For example, the method returns an
\texttt{ape::phylo} object if the NeXML document contains a single
phylogeny, an \texttt{ape::multiPhylo} object if it contains multiple
phylogenies arranged in a single \texttt{trees} block, and a list of
\texttt{ape::multiPhylo} objects if it contains multiple \texttt{trees}
blocks (the capability for which NeXML inherits from NEXUS).

If the location parameter with which the \texttt{nexml\_read()} method
is invoked is recognized as a URL, the method will automatically
download the document to the local working directory and read it from
there. This gives convenient and rapid access to phylogenetic data
published in NeXML format on the web, such as the content of the
phylogenetic data repository TreeBASE (Piel \emph{et al.} 2002, 2009).
For example, the following plots a tree in TreeBASE (using ape's plot
function):

\begin{Shaded}
\begin{Highlighting}[]
\NormalTok{tb_nex <-}\StringTok{ }\KeywordTok{nexml_read}\NormalTok{(}
\StringTok{"https://raw.github.com/TreeBASE/supertreebase/master/data/treebase/S100.xml"}\NormalTok{)}
\NormalTok{tb_phy <-}\StringTok{ }\KeywordTok{get_trees_list}\NormalTok{(tb_nex)}
\KeywordTok{plot}\NormalTok{(tb_phy[[}\DecValTok{1}\NormalTok{]]) }
\end{Highlighting}
\end{Shaded}

The method \texttt{get\_characters()} obtains character data matrices
from a \texttt{nexml} object, and returns them as a standard
\texttt{data.frame} R object with columns as characters and rows as
taxa:

\begin{Shaded}
\begin{Highlighting}[]
\NormalTok{nex <-}\StringTok{ }\KeywordTok{nexml_read}\NormalTok{(}\StringTok{"components/comp_analysis.xml"}\NormalTok{)}
\KeywordTok{get_characters}\NormalTok{(nex)}
\end{Highlighting}
\end{Shaded}

\begin{verbatim}
         log snout-vent length reef-dwelling
taxon_8             -3.2777799             0
taxon_9              2.0959433             1
taxon_10             3.1373971             0
taxon_1              4.7532824             1
taxon_2             -2.7624146             0
taxon_3              2.1049413             0
taxon_4             -4.9504770             0
taxon_5              1.2714718             1
taxon_6              6.2593966             1
taxon_7              0.9099634             1
\end{verbatim}

A NeXML data matrix can be of molecular (for molecular sequence
alignments), discrete (for most morphological character data), or
continuous type (for many trait data). To enable strict validation of
data types NeXML allows, and if their data types differ requires
multiple data matrices to be separated into different ``blocks''. Since
the \texttt{data.frame} data structure in R has no such constraints, the
\texttt{get\_characters()} method combines such blocks as separate
columns into a single \texttt{data.frame} object, provided they
correspond to the same taxa. Otherwise, a list of \texttt{data.frame}s
is returned, with list elements corresponding to characters blocks.
Similar to the methods for obtaining trees, there is also a method
\texttt{get\_characters\_list()}, which always returns a list of
\texttt{data.frame}s, one for each character block.

\subsection{Writing phylogenies and character
data}\label{writing-phylogenies-and-character-data}

The method \texttt{nexml\_write()} generates a NeXML file from its input
parameters. In its simplest invocation, the method writes a tree to a
file:

\begin{Shaded}
\begin{Highlighting}[]
\KeywordTok{data}\NormalTok{(bird.orders)}
\KeywordTok{nexml_write}\NormalTok{(bird.orders, }\DataTypeTok{file =} \StringTok{"birds.xml"}\NormalTok{)}
\end{Highlighting}
\end{Shaded}

The first argument to \texttt{nexml\_write()} is either an object of
type \texttt{nexml}, or any object that can be coerced to it, such as in
the above example an \texttt{ape::phylo} phylogeny. Alternatively,
passing a \texttt{multiPhylo} object would write a list of phylogenies
to the file.

In addition to trees, the \texttt{nexml\_write()} method also allows to
specify character data as another parameter. The following example uses
data from the comparative phylogenetics R package \texttt{geiger}
(Pennell \emph{et al.} 2014).

\begin{Shaded}
\begin{Highlighting}[]
\KeywordTok{library}\NormalTok{(}\StringTok{"geiger"}\NormalTok{)}
\KeywordTok{data}\NormalTok{(geospiza)}
\KeywordTok{nexml_write}\NormalTok{(}\DataTypeTok{trees =} \NormalTok{geospiza$phy, }
            \DataTypeTok{characters =} \NormalTok{geospiza$dat,}
            \DataTypeTok{file=}\StringTok{"geospiza.xml"}\NormalTok{)}
\end{Highlighting}
\end{Shaded}

Note that the NeXML format is well-suited for incomplete data: for
instance, here it does not assume the character matrix has data for
every tip in the tree.

\subsection{Validating NeXML}\label{validating-nexml}

File validation is a central feature of the NeXML format which ensures
that any properly implemented NeXML parser will always be able to read
the NeXML file. The function takes the path to any NeXML file and
returns \texttt{TRUE} to indicate a valid file, or \texttt{FALSE}
otherwise, along with a display of any error messages generated by the
validator.

\begin{Shaded}
\begin{Highlighting}[]
\KeywordTok{nexml_validate}\NormalTok{(}\StringTok{"geospiza.xml"}\NormalTok{)}
\end{Highlighting}
\end{Shaded}

\begin{verbatim}
[1] TRUE
\end{verbatim}

The \texttt{nexml\_validate()} function performs this validation using
the online NeXML validator (when a network connection is available),
which performs additional checks not expressed in the NeXML schema
itself (Vos \emph{et al.} 2012). If a network connection is not
available, the function falls back on the schema validation method from
the \texttt{XML} package (Lang 2013).

\subsection{Creating and populating \texttt{nexml}
objects}\label{creating-and-populating-nexml-objects}

Instead of packaging the various components for a NeXML file at the time
of writing the file, \texttt{RNeXML} also allows users to create and
iteratively populate in-memory \texttt{nexml} objects. The methods to do
this are \texttt{add\_characters()}, \texttt{add\_trees()}, and
\texttt{add\_meta()}, for adding characters, trees, and metadata,
respectively. Each of these functions will automatically create a new
nexml object if not supplied with an existing one as the last (optional)
argument.

For example, here we use \texttt{add\_trees()} to first create a
\texttt{nexml} object with the phylogeny data, and then add the
character data to it:

\begin{Shaded}
\begin{Highlighting}[]
\NormalTok{nexObj <-}\StringTok{ }\KeywordTok{add_trees}\NormalTok{(geospiza$phy)}
\NormalTok{nexObj <-}\StringTok{ }\KeywordTok{add_characters}\NormalTok{(geospiza$dat, nexObj)}
\end{Highlighting}
\end{Shaded}

The data with which a \texttt{nexml} object is populated need not share
the same OTUs. \texttt{RNeXML} automatically adds new, separate OTU
blocks into the NeXML file for each data matrix and tree that uses a
different set of OTUs.

Other than storage size, there is no limit to the number of phylogenies
and character matrices that can be included in a single NeXML document.
This allows, for example, to capture samples from a posterior
probability distribution of inferred or simulated phylogenies and
character states in a single NeXML file.

\subsection{Data documentation and annotation with built-in
metadata}\label{data-documentation-and-annotation-with-built-in-metadata}

NeXML allows attaching (``\emph{annotating}'') metadata to any data
element, and even to metadata themselves. Whether at the level of the
document as a whole or an individual data matrix or phylogeny, metadata
can provide bibliographic and provenance information, for example about
the study as part of which the phylogeny was generated or applied, which
data matrix and which methods were used to generate it. Metadata can
also be attached to very specific elements of the data, such as specific
traits, individual OTUs, nodes, or even edges of the phylogeny.

As described in Vos et al. (2012), to encode metadata annotations NeXML
uses the ``Resource Description Framework in Annotations'' (RDFa)
(Prud'hommeaux 2014). This standard provides for a strict
machine-readable format yet enables future backwards compatibility with
compliant NeXML parsers (and thus \texttt{RNeXML}), because the capacity
of a tool to \emph{parse} annotations is not predicated on
\emph{understanding} the meaning of annotations it has not seen before.

To lower the barriers to sharing well-documented phylogenetic data,
\texttt{RNeXML} aims to make recording useful and machine-readable
metadata easier at several levels.

First, when writing a NeXML file the package adds certain basic metadata
automatically if they are absent, using default values consistent with
recommended best practices (Cranston \emph{et al.} 2014). Currently,
this includes naming the software generating the NeXML, a time-stamp of
when a tree was produced, and an open data license. These are merely
default arguments to \texttt{add\_basic\_meta()} and can be configured.

Second, \texttt{RNeXML} provides a simple method, called
\texttt{add\_basic\_metadata()}, to set metadata attributes commonly
recommended for inclusion with data to be publicly archived or shared
(Cranston \emph{et al.} 2014). The currently accepted parameters include
\texttt{title}, \texttt{description}, \texttt{creator},
\texttt{pubdate}, \texttt{rights}, \texttt{publisher}, and
\texttt{citation}. Behind the scenes the method automatically anchors
these attributes in common vocabularies (such as Dublin Core).

Third, \texttt{RNeXML} integrates with the R package \texttt{taxize}
(Chamberlain \& Sz{ö}cs 2013) to mitigate one of the most common
obstacles to reuse of phylogenetic data, namely the misspellings and
inconsistent taxonomic naming with which OTU labels are often fraught.
The \texttt{taxize\_nexml()} method in \texttt{RNeXML} uses
\texttt{taxize} to match OTU labels against the NCBI database, and,
where a unique match is found, it annotates the respective OTU with the
matching NCBI identifier.

\subsection{Data annotation with custom
metadata}\label{data-annotation-with-custom-metadata}

The \texttt{RNeXML} interface described above for built-in metadata
allows users to create precise and semantically rich annotations without
confronting any of the complexity of namespaces and ontologies.
Nevertheless, advanced users may desire the explicit control over these
semantic tools that takes full advantage of the flexibility and
extensibility of the NeXML specification (Parr \emph{et al.} 2011; Vos
\emph{et al.} 2012). In this section we detail how to accomplish these
more complex uses in RNeXML.

Using a vocabulary or ontology terms rather than simple text strings to
describe data is crucial for allowing machines to not only parse but
also interpret and potentially reason over their semantics. To achieve
this benefit for custom metadata extensions, the user necessarily needs
to handle certain technical details from which the \texttt{RNeXML}
interface shields her otherwise, in particular the globally unique
identifiers (normally HTTP URIs) of metadata terms and vocabularies. To
be consistent with XML terminology, \texttt{RNeXML} calls vocabulary
URIs \emph{namespaces}, and their abbreviations \emph{prefixes}. For
example, the namespace for the Dublin Core Metadata Terms vocabulary is
``\url{http://purl.org/dc/elements/1.1/}''. Using its common
abbreviation ``dc'', a metadata property ``dc:title'' expands to the
identifier ``\url{http://purl.org/dc/elements/1.1/title}''. This URI
resolves to a human and machine-readable (depending on access)
definition of precisely what the term \texttt{title} in Dublin Core
means. In contrast, just using the text string ``title'' could also mean
the title of a person, a legal title, the verb title, etc. URI
identifiers of metadata vocabularies and terms are not mandated to
resolve, but if machines are to derive the maximum benefit from them,
they should resolve to a definition of their semantics in RDF.

\texttt{RNeXML} includes methods to obtain and manipulate metadata
properties, values, identifiers, and namespaces. The
\texttt{get\_namespaces()} method accepts a \texttt{nexml} object and
returns a named list of namespace prefixes and their corresponding
identifiers known to the object:

\begin{Shaded}
\begin{Highlighting}[]
\NormalTok{birds <-}\StringTok{ }\KeywordTok{nexml_read}\NormalTok{(}\StringTok{"birds.xml"}\NormalTok{)}
\NormalTok{prefixes <-}\StringTok{ }\KeywordTok{get_namespaces}\NormalTok{(birds)}
\NormalTok{prefixes[}\StringTok{"dc"}\NormalTok{]}
\end{Highlighting}
\end{Shaded}

\begin{verbatim}
                                dc 
"http://purl.org/dc/elements/1.1/" 
\end{verbatim}

The \texttt{get\_metadata()} method returns, as a named list, the
metadata annotations for a given \texttt{nexml} object at a given level,
with the whole NeXML document being the default level (\texttt{"all"}
extracts all metadata objects):

\begin{Shaded}
\begin{Highlighting}[]
\NormalTok{meta <-}\StringTok{ }\KeywordTok{get_metadata}\NormalTok{(birds) }
\NormalTok{otu_meta <-}\StringTok{ }\KeywordTok{get_metadata}\NormalTok{(birds, }\DataTypeTok{level=}\StringTok{"otu"}\NormalTok{)}
\end{Highlighting}
\end{Shaded}

The returned list does not include the data elements to which the
metadata are attached. Therefore, a different approach, documented in
the metadata vignette, is recommended for accessing the metadata
attached to data elements.

The \texttt{meta()} method creates a new metadata object from a property
name and content (value). For example, the following creates a
modification date metadata object, using a property in the PRISM
vocabulary:

\begin{Shaded}
\begin{Highlighting}[]
\NormalTok{modified <-}\StringTok{ }\KeywordTok{meta}\NormalTok{(}\DataTypeTok{property =} \StringTok{"prism:modificationDate"}\NormalTok{, }\DataTypeTok{content =} \StringTok{"2013-10-04"}\NormalTok{)}
\end{Highlighting}
\end{Shaded}

Metadata annotations in \texttt{NeXML} can be nested within another
annotation, which the \texttt{meta()} method accommodates by accepting a
parameter \texttt{children}, with the list of nested metadata objects
(which can themselves be nested) as value.

The \texttt{add\_meta()} function adds metadata objects as annotations
to a \texttt{nexml} object at a specified level, with the default level
being the NeXML document as a whole:

\begin{Shaded}
\begin{Highlighting}[]
\NormalTok{birds <-}\StringTok{ }\KeywordTok{add_meta}\NormalTok{(modified, birds) }
\end{Highlighting}
\end{Shaded}

If the prefix used by the metadata property is not among the built-in
ones (which can be obtained using \texttt{get\_namespaces()}), it has to
be provided along with its URI as the \texttt{namespaces} parameter. For
example, the following uses the
``\href{http://www.w3.org/TR/skos-reference/}{Simple Knowledge
Organization System}'' (SKOS) vocabulary to add a note to the trees in
the \texttt{nexml} object:

\begin{Shaded}
\begin{Highlighting}[]
\NormalTok{history <-}\StringTok{ }\KeywordTok{meta}\NormalTok{(}\DataTypeTok{property =} \StringTok{"skos:historyNote"}\NormalTok{,}
  \DataTypeTok{content =} \StringTok{"Mapped from the bird.orders data in the ape package using RNeXML"}\NormalTok{)}
\NormalTok{birds <-}\StringTok{ }\KeywordTok{add_meta}\NormalTok{(history, }
                \NormalTok{birds, }
                \DataTypeTok{level =} \StringTok{"trees"}\NormalTok{,}
                \DataTypeTok{namespaces =} \KeywordTok{c}\NormalTok{(}\DataTypeTok{skos =} \StringTok{"http://www.w3.org/2004/02/skos/core#"}\NormalTok{))}
\end{Highlighting}
\end{Shaded}

Alternatively, additional namespaces can also be added in batch using
the \texttt{add\_namespaces()} method.

By virtue of subsetting the S4 \texttt{nexml} object, \texttt{RNeXML}
also offers fine control of where a \texttt{meta} element is added, for
which the package vignette on S4 subsetting of \texttt{nexml} contains
examples.

Because NeXML expresses all metadata using the RDF standard, and stores
them compliant with RDFa, they can be extracted as an RDF graph,
queried, analyzed, and mashed up with other RDF data, local or on the
web, using a wealth of off-the-shelf tools for working with RDF (see
Prud'hommeaux (2014) or Hartig (2012)). Examples for these possibilities
are included in the \texttt{RNeXML} SPARQL vignette (a recursive acronym
for SPARQL Protocol and RDF Query Language, see
\url{http://www.w3.org/TR/rdf-sparql-query/}), and the package also
comes with a demonstration that can be run from R using the following
command: \texttt{demo("sparql", "RNeXML")}).

\subsection{Using metadata to extend the NeXML
standard}\label{using-metadata-to-extend-the-nexml-standard}

NeXML was designed to prevent the need for future non-interoperable
``flavors'' of the standard in response to new research directions. Its
solution to this inevitable problem is a highly flexible metadata system
without sacrificing strict validation of syntax and structure.

Here we illustrate how \texttt{RNeXML}'s interface to NeXML's metadata
system can be used to record and share a type of phylogenetic data not
taken into account when NeXML was designed, in this case stochastic
character maps (Huelsenbeck \emph{et al.} 2003). Such data assign
certain parts (corresponding to time) of each branch in a
time-calibrated phylogeny to a particular ``state'' (typically of a
morphological characteristic). The current de-facto format for sharing
stochastic character maps, created by \texttt{simmap} (Bollback 2006), a
widely used tool for creating such maps, is a non-interoperable
modification of the standard Newick tree format. This means that
computer programs designed to read Newick or NEXUS formats may fail when
trying to read in a phylogeny that includes \texttt{simmap} annotations.

In contrast, by allowing new data types to be added as --- sometimes
complex --- metadata annotations NeXML can accommodate data extensions
without compromise to its grammar and thus syntax In NeXML. To
illustrate how RNeXML facilitates extending the NeXML standard in this
way, we have implemented two functions in the package,
\texttt{nexml\_to\_simmap} and \texttt{simmap\_to\_nexml}. These
functions show how simmap data can be represented as \texttt{meta}
annotations on the branch length elements of a NeXML tree, and provide
routines to convert between this NeXML representation and the extended
\texttt{ape::phylo} representation of a \texttt{simmap} tree in R that
was introduced by Revell (2012). We encourage readers interested in this
capability to consult the example code in \texttt{simmap\_to\_nexml} to
see how this is implemented.

Extensions to NeXML must also be defined in the file's namespace in
order to valid. This provides a way to ensure that a URI providing
documentation of the extension is always included. Our examples here use
the prefix, \texttt{simmap}, to group the newly introduced metadata
properties in a vocabulary, for which the \texttt{add\_namespace()}
method can be used to give a URI as an identifier:

\begin{Shaded}
\begin{Highlighting}[]
\NormalTok{nex <-}\StringTok{ }\KeywordTok{add_namespaces}\NormalTok{(}\KeywordTok{c}\NormalTok{(}\DataTypeTok{simmap =} 
  \StringTok{"https://github.com/ropensci/RNeXML/tree/master/inst/simmap.md"}\NormalTok{))}
\end{Highlighting}
\end{Shaded}

Here the URI does not resolve to a fully machine-readable definition of
the terms and their semantics, but it can nonetheless be used to provide
at least a human-readable informal definition of the terms.

\subsection{Publishing NeXML files from
R}\label{publishing-nexml-files-from-r}

Data archiving is increasingly required by scientific journals,
including in evolutionary biology, ecology, and biodiversity (e.g.
Rausher et al. (2010)). The effort involved with preparing and
submitting properly annotated data to archives remains a notable barrier
to the broad adoption of data archiving and sharing as a normal part of
the scholarly publication workflow (Tenopir \emph{et al.} 2011; Stodden
2014). In particular, the majority of phylogenetic trees published in
the scholarly record are inaccessible or lost to the research community
(Drew \emph{et al.} 2013).

One of \texttt{RNeXML}'s aims is to promote the archival of
well-documented phylogenetic data in scientific data repositories, in
the form of NeXML files. To this end, the method
\texttt{nexml\_publish()} provides an API directly from within R that
allows data archival to become a step programmed into data management
scripts. Initially, the method supports the data repository Figshare
(\url{http://figshare.com}):

\begin{Shaded}
\begin{Highlighting}[]
\NormalTok{doi <-}\StringTok{ }\KeywordTok{nexml_publish}\NormalTok{(birds, }\DataTypeTok{repository=}\StringTok{"figshare"}\NormalTok{)}
\end{Highlighting}
\end{Shaded}

This method reserves a permanent identifier (DOI) on the figshare
repository that can later be made public through the figshare web
interface. This also acts as a secure backup of the data to a repository
and a way to share with collaborators prior to public release.

\section{Conclusions and future
directions}\label{conclusions-and-future-directions}

\texttt{RNeXML} allows R's ecosystem to read and write data in the NeXML
format through an interface that is no more involved than reading or
writing data from other phylogenetic data formats. It also carries
immediate benefits for its users compared to other formats. For example,
comparative analysis R packages and users frequently add their own
metadata annotations to the phylogenies they work with, such as
annotations of species, stochastic character maps, trait values, model
estimates and parameter values. \texttt{RNeXML} affords R the capability
to harness machine-readable semantics and an extensible metadata schema
to capture, preserve, and share these and other kinds of information,
all through an API instead of having to understand in detail the schema
underlying the NeXML standard. To assist users in meeting the rising bar
for best practices in data sharing in phylogenetic research (Cranston
\emph{et al.} 2014), \texttt{RNeXML} captures metadata information from
the R environment to the extent possible, and applies reasonable
defaults.

The goals for continued development of \texttt{RNeXML} revolve primarily
around better interoperability with other existing phylogenetic data
representations in R, such as those found in the \texttt{phylobase}
package (NESCENT R Hackathon Team 2014); and better integration of the
rich metadata semantics found in ontologies defined in the Web Ontology
Language (OWL), including programmatic access to machine reasoning with
such metadata.

\subsection{Acknowledgements}\label{acknowledgements}

This project was supported in part by the National Evolutionary
Synthesis Center (NESCent) (NSF \#EF-0905606), and grants from the
National Science Foundation (DBI-1306697) and the Alfred P Sloan
Foundation (Grant 2013-6-22). \texttt{RNeXML} started as a project idea
for the Google Summer of Code(TM), and we thank Kseniia Shumelchyk for
taking the first steps to implement it. We are grateful to F. Michonneau
for helpful comments on an earlier version of this manuscript, and
reviews by Matthew Pennell, Associate Editor Richard FitzJohn, and an
anonymous reviewer. At their behest, the reviews of FitzJohn and Pennell
can be found in this project's GitHub page at
\href{https://github.com/ropensci/RNeXML/issues/121}{github.com/ropensci/RNeXML/issues/120}
and
\href{https://github.com/ropensci/RNeXML/issues/120}{github.com/ropensci/RNeXML/issues/120},
together with our replies and a record of our revisions.

\subsection{Data Accessibility}\label{data-accessibility}

All software, scripts and data used in this paper can be found in the
permanent data archive Zenodo under the digital object identifier
\url{doi:10.5281/zenodo.13131} (Boettiger \emph{et al.} 2014). This DOI
corresponds to a snapshot of the GitHub repository at
\href{https://github.com/ropensci/RNeXML}{github.com/ropensci/RNeXML}.

\section*{References}\label{references}
\addcontentsline{toc}{section}{References}

Boettiger, C., Vos, R., Chamberlain, S. \& Lapp, H. (2014). RNeXML
v2.0.0. Retrieved from \url{http://dx.doi.org/10.5281/zenodo.13131}

Bollback, J. (2006).\emph{BMC Bioinformatics}, \textbf{7}, 88. Retrieved
from \url{http://dx.doi.org/10.1186/1471-2105-7-88}

Chamberlain, S.A. \& Sz{ö}cs, E. (2013). Taxize: Taxonomic search and
retrieval in r. \emph{F1000Research}. Retrieved from
\url{http://dx.doi.org/10.12688/f1000research.2-191.v2}

Cranston, K., Harmon, L.J., O'Leary, M.A. \& Lisle, C. (2014). Best
practices for data sharing in phylogenetic research. \emph{PLoS Curr}.
Retrieved from
\url{http://dx.doi.org/10.1371/currents.tol.bf01eff4a6b60ca4825c69293dc59645}

Drew, B.T., Gazis, R., Cabezas, P., Swithers, K.S., Deng, J., Rodriguez,
R., Katz, L.A., Crandall, K.A., Hibbett, D.S. \& Soltis, D.E. (2013).
Lost branches on the tree of life. \emph{PLoS Biol}, \textbf{11},
e1001636. Retrieved from
\url{http://dx.doi.org/10.1371/journal.pbio.1001636}

Hartig, O. (2012). An introduction to sPARQL and queries over linked
data. \emph{Web engineering} pp. 506--507. Springer Science + Business
Media. Retrieved from
\url{http://dx.doi.org/10.1007/978-3-642-31753-8_56}

Huelsenbeck, J.P., Nielsen, R. \& Bollback, J.P. (2003). Stochastic
mapping of morphological characters. \emph{Systematic Biology},
\textbf{52}, 131--158. Retrieved from
\url{http://dx.doi.org/10.1080/10635150390192780}

Lang, D.T. (2013). \emph{XML: Tools for parsing and generating xML
within r and s-plus.} Retrieved from
\url{http://CRAN.R-project.org/package=XML}

Maddison, D., Swofford, D. \& Maddison, W. (1997). NEXUS: An extensible
file format for systematic information. \emph{Syst. Biol.}, \textbf{46},
590--621. Retrieved from
\url{http://www.ncbi.nlm.nih.gov/pubmed/11975335}

Midford, P., Dececchi, T., Balhoff, J., Dahdul, W., Ibrahim, N., Lapp,
H., Lundberg, J., Mabee, P., Sereno, P., Westerfield, M., Vision, T. \&
Blackburn, D. (2013). The vertebrate taxonomy ontology: A framework for
reasoning across model organism and species phenotypes. \emph{J. Biomed.
Semantics}, \textbf{4}, 34. Retrieved from
\url{http://dx.doi.org/10.1186/2041-1480-4-34}

NESCENT R Hackathon Team. (2014). \emph{Phylobase: Base package for
phylogenetic structures and comparative data}. Retrieved from
\url{http://CRAN.R-project.org/package=phylobase}

O'Meara, B. (2014). CRAN task view: Phylogenetics, especially
comparative methods. Retrieved from
\url{http://cran.r-project.org/web/views/Phylogenetics.html}

Paradis, E., Claude, J. \& Strimmer, K. (2004). APE: Analyses of
phylogenetics and evolution in R language. \emph{Bioinformatics},
\textbf{20}, 289--290.

Parr, C.S., Guralnick, R., Cellinese, N. \& Page, R.D.M. (2011).
Evolutionary informatics: unifying knowledge about the diversity of
life. \emph{Trends in ecology \& evolution}, \textbf{27}, 94--103.
Retrieved from \url{http://www.ncbi.nlm.nih.gov/pubmed/22154516}

Pennell, M.W., Eastman, J.M., Slater, G.J., Brown, J.W., Uyeda, J.C.,
Fitzjohn, R.G., Alfaro, M.E. \& Harmon, L.J. (2014). Geiger v2.0: An
expanded suite of methods for fitting macroevolutionary models to
phylogenetic trees. \emph{Bioinformatics}, \textbf{30}, 2216--2218.

Piel, W.H., Chan, L., Dominus, M.J., Ruan, J., Vos, R.A. \& Tannen, V.
(2009). TreeBASE v. 2: A database of phylogenetic knowledge. Retrieved
from \url{http://www.e-biosphere09.org}

Piel, W.H., Donoghue, M.J. \& Sanderson, M.J. (2002). TreeBASE: A
database of phylogenetic information. \emph{The interoperable `catalog
of life'} (eds J. Shimura, K.L. Wilson \& D. Gordon), pp. 41--47.
Research report. National Institute for Environmental Studies, Tsukuba,
Japan. Retrieved from
\url{http://donoghuelab.yale.edu/sites/default/files/124_piel_shimura02.pdf}

Prud'hommeaux, E. (2014). SPARQL query language for rDF. \emph{W3C}.
Retrieved from \url{http://www.w3.org/TR/rdf-sparql-query/}

R Core Team. (2014). \emph{R: A language and environment for statistical
computing}. R Foundation for Statistical Computing, Vienna, Austria.
Retrieved from \url{http://www.R-project.org/}

Rausher, M.D., McPeek, M.A., Moore, A.J., Rieseberg, L. \& Whitlock,
M.C. (2010). Data archiving. \emph{Evolution}, \textbf{64}, 603--604.
Retrieved from \url{http://dx.doi.org/10.1111/j.1558-5646.2009.00940.x}

Revell, L.J. (2012). Phytools: An r package for phylogenetic comparative
biology (and other things). \emph{Methods in Ecology and Evolution},
\textbf{3}, 217--223.

Stodden, V. (2014). The scientific method in practice: Reproducibility
in the computational sciences. \emph{SSRN Journal}. Retrieved from
\url{http://dx.doi.org/10.2139/ssrn.1550193}

Stoltzfus, A., O'Meara, B., Whitacre, J., Mounce, R., Gillespie, E.L.,
Kumar, S., Rosauer, D.F. \& Vos, R.A. (2012). Sharing and re-use of
phylogenetic trees (and associated data) to facilitate synthesis.
\emph{BMC Research Notes}, \textbf{5}, 574. Retrieved from
\url{http://dx.doi.org/10.1186/1756-0500-5-574}

Tenopir, C., Allard, S., Douglass, K., Aydinoglu, A.U., Wu, L., Read,
E., Manoff, M. \& Frame, M. (2011). Data sharing by scientists:
Practices and perceptions (C. Neylon, Ed.). \emph{PLoS ONE}, \textbf{6},
e21101. Retrieved from
\url{http://dx.doi.org/10.1371/journal.pone.0021101}

Vos, R.A., Balhoff, J.P., Caravas, J.A., Holder, M.T., Lapp, H.,
Maddison, W.P., Midford, P.E., Priyam, A., Sukumaran, J., Xia, X. \&
Stoltzfus, A. (2012). NeXML: Rich, extensible, and verifiable
representation of comparative data and metadata. \emph{Systematic
Biology}, \textbf{61}, 675--689. Retrieved from
\url{http://dx.doi.org/10.1093/sysbio/sys025}

\end{document}